\documentclass{articlek}
\usepackage{amsmath}

\renewcommand{\refname}{REFERENCES}
\def\be{\begin{equation}}
\def\ee{\end{equation}}

\def\BibTeX{{\rm B\kern-.05em{\sc i\kern-.025em b}\kern-.08em
            T\kern-.1667em\lower.7ex\hbox{E}\kern-.125emX}}

\usepackage{graphicx}
\begin{document}
\sloppy
\twocolumn[{
{\large\bf ULTRARELATIVISTIC HEAVY ION COLLISIONS: EXPLORING THE
PHASE DIAGRAM OF QCD}\\

{\small B. Tom\'a\v{s}ik, tomasik@fpv.umb.sk,\\
Fakulta pr\'irodn\'ych vied, Univerzita Mateja Bela, Tajovsk\'eho 40,
97401 Bansk\'a Bystrica, Slovakia, and \\
\'Ustav vedy a v\'yskumu, Univerzita Mateja Bela, Cesta na amfite\'ater 1,
97401 Bansk\'a Bystrica, Slovakia}\\


{\bf ABSTRACT.} I present the motivation for studying nuclear
collisions at ultrarelativistic energies which is to map the phase
diagram of strongly interacting matter under very extreme
conditions. The relevant experimental efforts are overviewed and
problems in connecting observables with physics of the hot and
dense matter are pointed out. I review the current highlights of
the experimental programme: excitation functions of various
quantities from the SPS which suggest a possibility of the onset
of deconfinement, and the evidence of deconfined perfect fluid at
RHIC. \vspace*{5ex} }]

\section{THE PHASE DIAGRAM OF QCD}
\label{phasdiag}

Phase diagram is graphical representation of the qualitative
properties of {\em bulk} matter and of the conditions under which
these properties change. Microscopically, the bulk properties
result from the underlying interaction between constituents of the
material. This also gives typical energy or temperature scales at
which phase transitions would happen.

All properties of materials we usually deal with are due to
electromagnetic interaction. Let us take water as an example.
Boiling and condensing is connected with rearranging bonds between
the molecules, so intermolecular van der Waals interaction gives
the typical energy scale, which is of order $10^{-2}$~eV per
molecule. By using typical molecular densities one can get an
estimate for critical energy density and temperature.

At somewhat higher energy the electrically neutral atoms break up
into positively charged ions and electrons. This results in plasma
phase. The typical ionisation energy of an atom is of the order of
electronvolt, thus about a factor 100 higher temperature than for
boiling would be expected.

We can ask if the strong interaction leads to analogic phase
transitions. Strong interaction acts on entities with colour
charge; the elementary participants are quarks and gluons.
Similarly to electromagnetic interaction which acts also between
electrically neutral atoms and molecules via van der Waals forces,
there is strong interaction between colourless hadrons, e.g.\
nucleons within a nucleus. Nuclear matter is held together by
strong interaction.

The typical energy scale of strong interaction between nucleons in
a nucleus shows up as the binding energy. In analogy to phase
transition from liquid water to vapor known from everyday life, it
turns out that there is a first-order ``liquid--gas'' phase
transition in nuclear matter \cite{siemens,panag}. The typical
temperature is of order of few MeV\footnote{Here I use natural
system of units where Boltzmann constant $k_B=1$ as well as $\hbar
= c = 1$. Temperature in SI units is obtained by expressing $T$
first in Joules and then dividing by $k_B$.}.

Is there a strong interaction analogy to electromagnetic plasma
phase with free charges? This appears more problematic: while
electric charges can exist alone in vacuum, colour charges {\em
cannot}. They always must be bound together within a colourless
object. This property of quarks as carriers of colour charge is
called {\em confinement}. Nevertheless, if the energy density is
large enough, loosely speaking if the colour charges are packed
together very densely, the quarks may not be anymore bound to
hadrons but can freely roam over the whole region occupied by all
quarks together \cite{colper}. Such a system is labeled {\em
``quark-gluon plasma'' (QGP)}.

\begin{figure}[ht]
\centerline{\includegraphics[width=80mm]{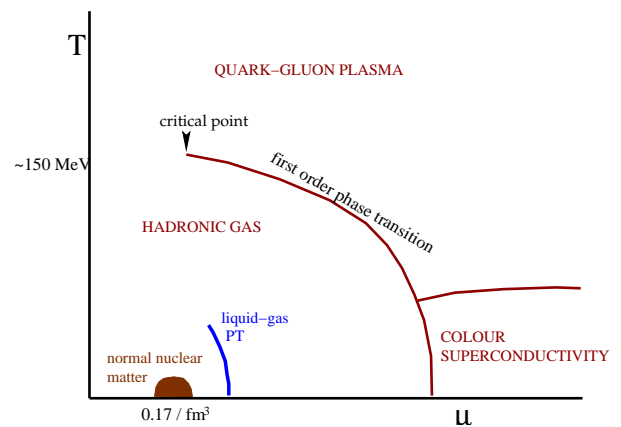}}
\vspace{-2mm}
\caption{%
Schematic view of the phase diagram of nuclear matter as a
function of baryochemical potential and the temperature. Note that
there are no scales put on the axes. \label{phd}}
\end{figure}
In principle, the phase diagram is determined by the Lagrangian of
underlying interaction. However, in case of the QCD Lagrangian it
is technically too complicated to calculate from the first
principles, yet. Its structure (Fig.~\ref{phd}) is qualitatively
and to certain level quantitatively known from effective theories
and numerical simulations. The phase diagram is shown as
characterised by the temperature $T$ and baryochemical potential.
Baryochemical potential is a measure of {\em net} baryon density.
By changing temperature not only typical momentum of quanta varies
but also the densities of baryons {\em and} antibaryons. We must
realise that the typical energy scales in this phase diagram are
large, and relativistic effects of particle creation may appear.
Thus the particle numbers are not conserved; what is conserved are
quantum numbers like baryon number.

At moderately high net baryon density there is the line of
first-order liquid-gas phase transition. Farther to asymptotically
high baryochemical potential while still at rather low temperature
we find the regime of colour superconductivity \cite{sc}. This is
an analogue to regular superconductivity which appears if there is
arbitrarily weak {\em attractive} interaction between fermions on
Fermi surface. Here, that role is played by certain channels of
strong interaction. There are several different phases of colour
superconductor, but we shall not go in such details here. Due to
the low temperature required, such phase is not produced in
nuclear collisions. It might be present in neutron stars, as they
are rather cool: only some billion Kelvin. (If you think that this
is hot, read below.) Still, there is no clear observational
evidence for colour superconductor in a neutron star so far
\cite{nstars}.

\begin{figure}[ht]
\centerline{\includegraphics[width=80mm]{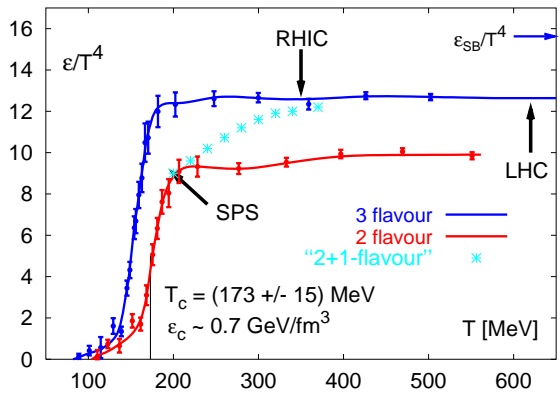}}
\vspace{-2mm}
\caption{%
The energy density normalised by $T^4$ as a function of temperature
from lattice QCD. Simulations with 2 and 3 light quark flavours and with
2 light and one heavy flavour are presented. Labels show the energy densities
reached at various accelerators.  Figure from \cite{karsch}.
\label{f:latt}}
\end{figure}

The regime I want to focus on is that of high temperature. We see
in Fig.~\ref{phd} that at high temperature the plasma phase is
expected. The equation of state at vanishing {\em net} baryon
density can be calculated numerically in so-called {\em lattice
QCD} approach \cite{karsch}. Such calculations indicate that at
temperature $173\pm 15$~MeV the energy density changes as a
function of temperature dramatically but smoothly (Fig.\ \ref{f:latt}).
In SI units this temperature is about $2\cdot 10^{12}$~K, which is
about 100,000 times hotter than in the centre of the Sun! Two
phase transitions happen here: confinement vanishes and chiral
symmetry is restored. The former is observed in the simulation as
a change of quark-quark potential from linear rise to an
asymptotically constant value at large distances; the latter as
vanishing of vacuum expectation value of chiral condensate
$\langle \bar \psi \psi \rangle$.

At large enough non-zero net baryon density  first order phase
transition appears. The critical line ends in a critical point of
second-order phase transition. Exact position of this point is of
fundamental interest, although so far there are only calculations
based on effective theories or rather unreliable estimates from
lattice QCD \cite{stephanov,foka}.

\section{CREATING EXTREME CONDITIONS}
\label{ex}

Have such extreme conditions ever been realised in nature? Can we create
them artificially? The answer to both questions is positive.

They existed in  early Universe which evolved from a very hot and
dense beginning through continuous expanding and cooling. The
relation between time and temperature in the early phase was
\begin{equation}
\frac{1}{t^2} = g_* \frac{4\pi^4}{90}
\frac{k_B^4 T^4}{\hbar^2 c^4 M_{\rm Pl}^2}\, ,
\end{equation}
where $M_{\rm Pl} = \sqrt{\hbar c/G_N}$ is the Planck mass and
$g_*$ is the effective number of degrees of freedom (loosely
speaking, how many different particle species contribute to the
energy density). According to this relation, temperature around
150--200~MeV was reached at  time 10--100 microseconds after the
Big Bang. This is the time when hadrons were born.

The amazing thing is that we can re-create a small ``early
Universe'' in the lab and study ``Little Bangs''! This is done
with the help of ultrarelativistic nuclear collisions. By
colliding large nuclei energy is converted into new quanta with
large typical momentum. At the same time the system is large
enough so that we can have bulk matter consisting of many
interacting constituents. At sufficient collision energy the
energy density within the system will reach up to the regime of
QGP.

We have to realise that the colliding nuclei contain only baryons
and no antibaryons, thus always matter with positive net baryon
density is created. The higher the collision energy, the more
secondary particles can be created. Among created particles there
are as many baryons as antibaryons and so the ratio of baryons to
antibaryons becomes smaller at higher collision energies.
Consequently, the baryochemical potential decreases.

\begin{figure}[bh]
\centerline{\includegraphics[width=80mm]{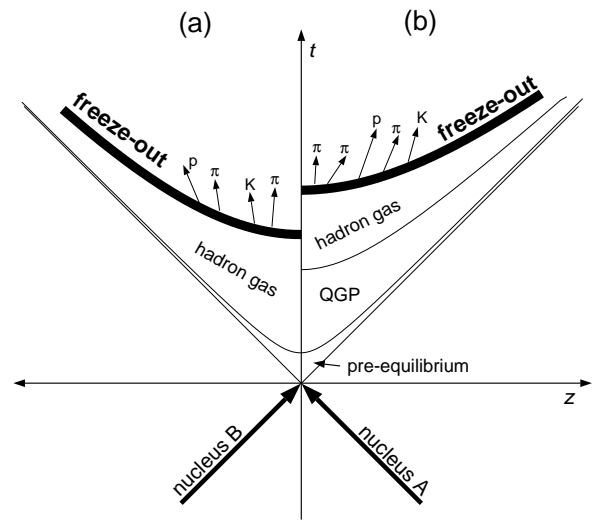}}
\vspace{-2mm}
\caption{%
The space-time diagram of longitudinal coordinate and time of the
evolution of fireball (a) without and (b) with the production of
quark-gluon plasma. \label{scen}}
\end{figure}

Hence, nuclear collisions at various energies can map various
regions of the phase diagram. Unfortunately, such collisions
create rapidly evolving systems---so-called fireballs---which
makes it rather complicated to study the generated matter. Let us
have a flash view of the evolution of fireball (Fig.~\ref{scen}).
\begin{enumerate}
\item
Colliding nuclei have relativistic energy, so before the collision
they appear in the centre of mass system Lorentz contracted. The
factor $\gamma$ spans from about 4 (at lowest SPS energy) up to
100 at RHIC and even more than 2000 at the LHC.
\item
If two nuclei collide at high energy, the incident nucleons and/or
partons tend to continue their longitudinal motion while being
just slowed down somewhat. After they have passed through each
other, the energy they have lost converts into new quanta in the
region between the receding original nucleons/partons. Just after
the collision the energy density of this system is maximum and the
matter could find itself in the deconfined phase, if it is high
enough.
\item
The longitudinal motion continues and thus the system expands
longitudinally. Consequently, it cools down. Due to inner pressure
also transverse expansion builds up which accelerates the cooling
process. Hence, even if the matter was in QGP phase initially,
eventually it changes to hadronic phase. This may happen via
first-order phase transition in the part of the phase diagram at
higher baryochemical potential or via
 rapid crossover.
\item
In any case, the fireball thus ends up in a hadronic phase where is still
expands very fast.
\item
Eventually, the density of the system becomes so low that no more
rescattering between its constituents appear and momenta of
hadrons freeze-out. These are the momenta observed in detectors.
It is important to realise that {\em hadrons} are observed and not
quarks or gluons.
\end{enumerate}
This scenario runs very quickly. From the initial impact until the
freeze-out it lasts about 10 fm/$c$ or more\footnote{The time unit
fm/$c$ is given by the time it takes for the light to pass the
typical size of proton, which is 1 femtometre (called also fermi).
This is the typical time scale in high energy physics. In SI it
corresponds to about $3\cdot 10^{-24}$~s.}. Unfortunately, we are
most interested in studying the hottest and densest phase of
matter, but we observe hadrons escaping from the less excited
hadronic gas. The main technical problem in this field is thus
relating observable quantities with the properties of early hot
state of matter.

Before addressing this problem, let me list the existing, recent,
and forthcoming experimental facilities devoted to this kind of
experiments. For better overview, they are listed in
Table~\ref{t-exp}.
\begin{table}
\caption{ Overview of accelerators used for ultrarelativistic
heavy ion studies. I list the maximum energy per participating
nucleon in the nucleon-nucleon centre of mass system and whether
the setup is fix target experiments or it is a collider.
\label{t-exp}}
\begin{center}
\begin{tabular}{|cccc|}
\hline\hline machine & lab & $\sqrt{s_{NN}^{\rm max}}$ & target or
collider\\ \hline\hline \multicolumn{4}{|l|}{recent} \\ \hline AGS
& BNL & 4.8~GeV & target \\ SPS & CERN & 17 GeV & target \\
\hline\hline \multicolumn{4}{|l|}{present} \\ \hline RHIC & BNL&
200~GeV & collider \\ \hline\hline
\multicolumn{4}{|l|}{forthcoming}\\ \hline LHC & CERN & 5500~GeV &
collider\\ FAIR & GSI & 8~GeV & target\\ \hline\hline
\end{tabular}
\end{center}
\end{table}
Baseline in many observed effects has been provided by the
Alternating Gradient Synchrotron (AGS) of Brookhaven National
Laboratory (BNL). This machine is currently used as preaccelerator
for RHIC (to be mentioned later). Up until recently, experiments
have been performed at CERN's Super Proton Synchrotron (SPS) which
accelerated nuclei up to the size of lead. These brought
interesting results which could indicate the onset of quark-gluon
plasma production in this energy region. Currently, world's
strongest heavy ion accelerator is Relativistic Heavy Ion Collider
(RHIC), where strongly interacting QGP behaving like a perfect
fluid has been discovered. In less than a year from now the Large
Hadron Collider (LHC) at CERN will be commissioned. Heavy ion
studies---in particular with the dedicated detector ALICE---are
planned at this machine among others. In 2014 completion of the
Facility for Antiproton and Ion Research (FAIR) of the
Gesellschaft f\"ur Schwerionenforschung (GSI) in Darmstadt is
scheduled, which will deliver high luminosity ion beams with
energies up to 34 GeV per nucleon for fixed target experiments.

In the following we shall  mainly talk about results from
experiments performed with lead or gold beams. There were
experiments with smaller nuclei, as well (In, Cu, S, O, \dots) in
order to gain better systematics of various observables.


\section{FREEZE-OUT STATE: THE GROSS PICTURE}
\label{fos}

The early hot phase of the collision is not observable directly.
There are basically two approaches which can be used in order to
learn about it.
\begin{enumerate}
\item
By studying momentum distributions of low $p_t$ hadrons one can
extract information about the state of the fireball at the time of
its breakup. This state can be reconstructed and one can deduce
the evolution of the fireball which could have lead to it. At
least, this method can discriminate theoretical scenarios which
lead to the wrong final state.
\item
Probes which are produced early and manage to escape the fireball
carry direct information about the hot matter. These are e.g.\
particles which interact only electromagnetically, like photons
and (di)leptons. Recently, hard jets have provided very clean
probe of the matter, which we shall discuss below.
\end{enumerate}
Let us first have a look at what hadrons tell us about the final state.

\begin{figure}[ht]
\centerline{\includegraphics[width=80mm]{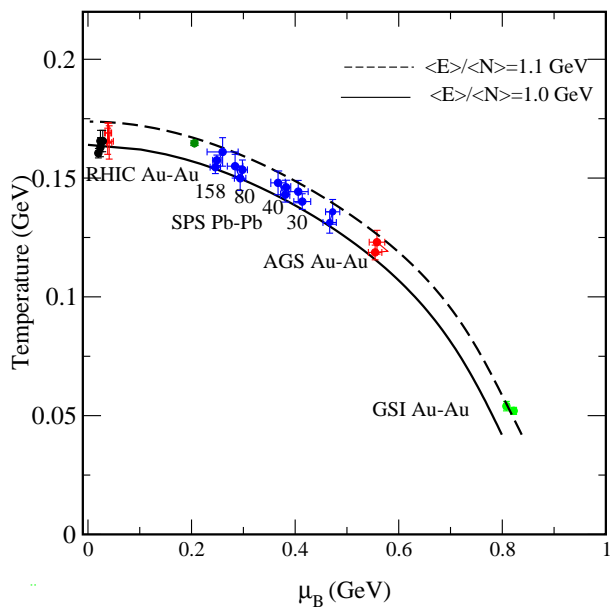}}
\vspace{-2mm}
\caption{%
Temperatures and baryochemical potentials extracted from fits to
chemical compositions of hadrons produced in nuclear collisions at
various energies. Figure taken from \cite{clre}.\label{f-cley}}
\end{figure}

It is interesting to analyse their chemical composition. This
means that we study how many hadrons of different sorts are there
among the observed particles. Interestingly, it turns out that the
chemical composition can be very reasonably described by chemical
equilibrium characterised by some value of temperature and
baryochemical potential. Moreover, the values of temperature and
chemical potential inferred for all collision energies lay on a
well defined curve in the $T$--$\mu$ plane (Fig.~\ref{f-cley})
\cite{clre}.
Results for RHIC and higher SPS energies seem to coincide with the
region of rapid crossover from hadronic to partonic system and
thus suggest that the abundances are fixed through the process of
hadronisation. Chemical freeze-out at lower SPS energies, about 6
to 10 GeV per nucleon in the centre of mass system, might be close
to the critical point of the phase diagram, though the precise
position of this point is unknown, yet \cite{foka}. This gives the
first hint about regions of phase diagram which can be explored
using different facilities: RHIC and LHC will look at hot matter
at small baryochemical potential likely undergoing a smooth
crossover to hadronic phase. On the other hand, SPS and the future
FAIR can search for the line of first order phase transition and
the critical point.

From single-particle momentum spectra and momentum correlations we
know that the fireball expands transversely rather fast
\cite{twh,rhicexp}. In a transversely expanding fireball
transverse momentum spectra are blueshifted due to Doppler effect.
The inverse exponential slope in non-relativistic approximation is
given as \cite{CL96}
\begin{equation}
T_* = T + m\langle v_t \rangle^2 \, , \quad \mbox{where} \,\,
\frac{dN}{p_t\, dp_t} \propto \exp\left ( -\frac{p_t}{T_*}\right
)\, .
\end{equation}
Thus from spectra of several identified species (with different
masses) we can extract the average transverse collective expansion
velocity $\langle v_t\rangle$. In central collisions at RHIC it
reaches up to 0.6$c$ \cite{rhicexp}.

This figure is also supported by momentum correlations. They
measure primarily the size of the emitting region. With rather
small variation with collision energy in central Pb+Pb and Au+Au
collisions the emitting fireball is about twice as big
transversely as the original nucleus! The average momentum
dependence of correlation functions is also consistent with strong
transverse expansion \cite{myrev,lisa}. Conclusion is that we do
observe {\em collectively behaving matter} and not just a bunch of
independent nucleon-nucleon collisions and that there is strong
pressure in the early phase of the collision.

\section{PERFECT FLUID AT RHIC}
\label{perfluid}

Interesting results have been observed in non-central collisions.
Here, the two nuclei collide so that the reaction zone is not
azimuthally symmetric but rather has elliptic shape with shorter
size in direction of the impact parameter. This initial shape
anisotropy is translated to an anisotropy in transverse momentum
spectra: more particles are produced in the direction of impact
parameter. This is quantitatively expressed by Fourier
decomposition of the spectrum
\begin{multline}
\frac{d^3N}{p_t\, dp_t\, dy\, d\phi} =
\frac{1}{2\pi}\, \frac{d^2N}{p_t\, dp_t\, dy} \\
\times
\left ( 1 + 2v_2(p_t,y) \cos(\phi-\phi_r)
+  \dots\right )
\end{multline}
(other terms of the decomposition vanish by symmetry at
midrapidity, $\phi_r$ is the azimuthal angle of impact parameter).
The second-order coefficient $v_2$ is called {\em elliptic flow}.

Momentum anisotropy measured by the elliptic flow is very
sensitive probe of conditions in the early hot phase. Data
indicate that more transverse flow is produced in the direction of
impact parameter than perpendicularly to it. This has natural
explanation in hydrodynamics. Flow results from pressure
gradients. The smaller size of the reaction zone in the direction
of impact parameter implies larger pressure gradients and more
flow. This mechanism of generating flow anisotropy vanishes with
time as the fireball becomes azimuthally symmetric through
expansion and so large flow anisotropy must be generated in the
earliest phase.

At RHIC, ideal hydrodynamics describes the transverse expansion
and transverse momentum spectra (almost) perfectly \cite{HKqgp3}.
(It only breaks for about 1\% of all produced particles with high
$p_t$.) This has not been so at lower collision energies like SPS.
Technically, hydrodynamics consists of solving conservation laws
and the equation of state. The simulation has been tuned on
description of central collisions, i.e., the prescription for
determining initial conditions has been fixed there in such a way
that spectra were reproduced. Then, without any other tuning
elliptic flow in non-central collisions has been successfully
predicted, as well \cite{HKqgp3}. It should be stressed that the
model was successful {\em only} if QGP phase was assumed during
the early hot period of fireball evolution!

Recall that data were reproduced with {\em ideal} hydrodynamics,
i.e.\ with vanishing viscosity. Different simulations with a
kinetic model have shown that the mean free path must be extremely
short, otherwise the simulation would fall below the measured
$v_2$ \cite{molnar}. Recall that in a kinetic approach shear
viscosity grows with the mean free path because momentum can be
transferred to larger distances. It has also been shown that
``viscous'' corrections of the elliptic flow just due to
modifications of locally thermal distribution are large and always
decrease $v_2$ \cite{teaney}.
The conclusion is that the dimensionless\footnote{%
This ratio is dimensionless in natural units. In SI, dimension of
viscosity is $[\eta]=\mbox{N}\cdot \mbox{s}\cdot \mbox{m}^{-2}$
and that of entropy density is  $[s] = \mbox{J}\cdot
\mbox{K}^{-1}\cdot \mbox{m}^{-3}$, so $[\eta/s] = \mbox{K}\cdot
\mbox{s} = [\hbar/k_B]$. } ratio of shear viscosity to entropy
density $\eta/s$ must be very small. Such a small viscosity is
consistent with the picture of plasma with {\em strongly
interacting} colour charges \cite{shuryak}, because weakly
interacting gas would have long mean free path and larger
viscosity. Calculations indicate that just above the critical
temperature the value of $\eta/s$ is as small 0.2
\cite{Nakamura:2004sy,Peshier:2005pp}. For comparison, the
corresponding value for water at normal conditions would be of
order\footnote{%
I should say that at higher temperature $\eta/s$ of water can be
as small as about 2, but even this is still much bigger than
QGP.}%
$10^3$. In fact, the strongly interacting quark-gluon plasma
is the most perfect liquid known.


\section{DECONFINEMENT AT RHIC}
\label{deco}

\begin{figure}[t]
\centerline{\includegraphics[width=80mm]{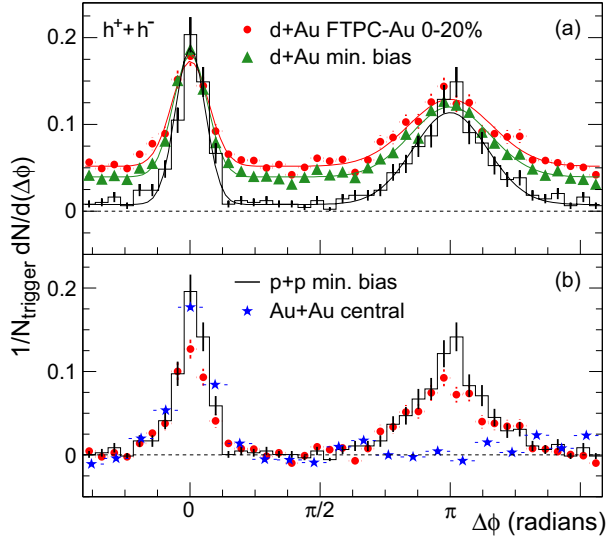}}
\vspace{-2mm}
\caption{%
Azimuthal correlation function of high $p_t$ particles in pp,
d+Au, and Au+Au collisions at $\sqrt{s} = 200 \, A\mbox{GeV}$.
Angle 0 is oriented in direction of a high $p_t$ jet trigger
particle. Figure taken from \cite{STARjets}.\label{f-jquench}}
\end{figure}
The previous section indicated that matter produced at RHIC is deconfined.
There is another spectacular demonstration of this: the jet quenching.

Jets are well known in high-energy collisions of simpler systems.
For example, in proton-proton collisions two partons from incoming
protons can exchange large momentum and secondary particles are
produced in two rather focused showers around the leading struck
out partons. Due to momentum conservation, jets can never be
produced alone. Most often, there is one associated jet in the
opposite direction, though in some cases three or more associated
jets can appear and rarely even counterbalancing photon or lepton
is emitted.

The STAR collaboration at RHIC has studied azimuthal correlations
of high $p_t$ particles associated with jets
(Fig.~\ref{f-jquench}) \cite{STARjets}. In proton-proton and
deuteron-gold collisions they found indeed that jets were produced
in pairs separated by $180^\circ$ in azimuthal angle. However, in
central Au+Au collisions the associated jet vanished (stars in
Fig.~\ref{f-jquench}). Since momentum conservation still holds,
the associated jet must have been produced but the leading
particle has been completely slowed down when traversing the
matter of the fireball. This is a signature of large energy loss
of leading jet parton which is only possible in a medium with high
density of colour charge, such as QGP.

Recently, more detailed studies showed that jets with very large
transverse momentum poke out through the medium \cite{Magestro:2005vm}.
This opens possibility to study the energy loss of leading partons
quantitatively and make conclusions about the density of early
hot phase.


\section{THE ONSET OF DECONFINEMENT}
\label{ondeco}

If QGP is produced at RHIC, what is the {\em minimum} collision
energy at which deconfinement sets in? In order to answer this
question one would study excitation functions, i.e.\ dependences
of various quantities on the collision energy.

\begin{figure}[ht]
\centerline{\includegraphics[width=80mm]{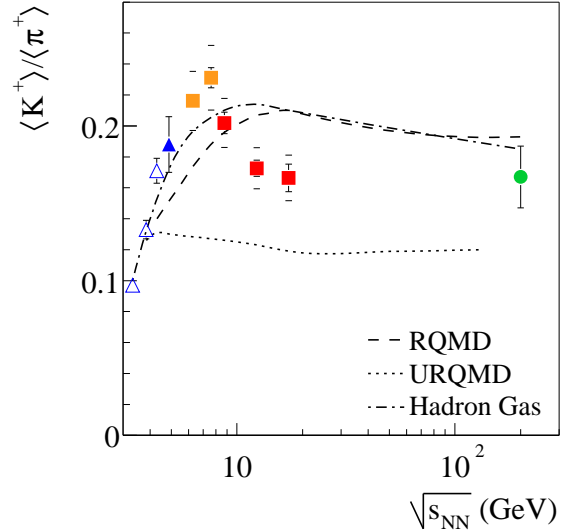}}
\vspace{-5mm}
\caption{%
Excitation function of the ratio of multiplicities of positively
charged kaons to pions. Curves present results of various model
calculations. Data of the NA49 Collaboration, figure taken from
\cite{Blume:2004vq}. \label{f:excit}}
\end{figure}

One example is shown in Fig.~\ref{f:excit} \cite{Blume:2004vq}.
The ratio of multiplicities of positively charged kaons to pions
first rapidly grows as a function of collision energy. It peaks
around $\sqrt{s}\approx 7\, A\mbox{GeV}$ drops down a little and
then it levels off. This interesting non-monotonic
structure---which has been dubbed ``the horn''---gives a first
hint that something may be happening here. In addition to this, in
the same energy region where the horn of Fig.~\ref{f:excit} drops
to the right one observes a plateau in the excitation function of
the mean $p_t$ of kaons. Finally, the number of produced pions per
participating nucleon changes its slope as a function of collision
energy just at the same place as the horn \cite{Blume:2004vq}.

These observations are successfully interpreted in framework of
so-called Statistical Model of Early Stage \cite{GG} which assumes
the onset of deconfinement where the horn appears. However, if
indeed QGP is produced here, then no hadronic model should be able
to describe all data. Thus it is important to play {\em Advocatus
Diaboli} and test hadronic models by comparing their predictions
with data. Indeed, most of them fail in reproducing these
excitation functions \cite{rqmdhsd,mosel,toneev}. Recently, we
constructed hadronic kinetic model \cite{toko} which reproduced
the horn of Fig.~\ref{f:excit}. It remains to test this model on
other pieces of data before it can be confirmed or ruled out.

\section{CONCLUSIONS}
\label{conc}

Let me summarise my selection of  current highlights of
ultrarelativistic heavy ion programme.
\begin{enumerate}
\item
QGP may be produced at lowest energies studied at the SPS. Careful
investigations, also employing the new FAIR machine, are necessary
in order to confirm this.
\item
\label{dva} Jet quenching at RHIC implies that we have matter
which eats jets. The only known medium able to do this is
quark-gluon plasma.
\item
\label{tri} Elliptic flow at RHIC is as large as it can possibly
be in perfect fluid and so the viscosity of QGP is tremendously
small. This is consistent with the regime of strongly interacting
QGP at temperatures just above $T_c$.
\end{enumerate}

The Large Hadron Collider to be commissioned next year will shift
the frontiers in points \ref{dva} and \ref{tri}. Due to higher
incident energy, jets which probe the early hot matter will be
produced copiously. At the larger initial temperature the medium
could find itself in a not-so-strongly interacting regime as the
strong coupling decreases for interactions at large momentum
scale. This would lead to larger viscosity and it would be
interesting to observe a corresponding decrease of elliptic flow.
Such an observation would nicely extend our knowledge of the QCD
phase diagram.


\end{document}